\documentclass[12pt]{iopart}
\usepackage{iopams}  
\usepackage{graphicx}

\usepackage{bm}

\def\KeyWord#1{$\backslash$\IfColor{$\!\!$\textRed{#1}\textBlack}{#1}$\!\!$}

\begin{document}
\title{Large number of receptors may reduce cellular response time variation}

\author{Xiang Cheng, Lina Merchan, Martin Tchernookov} 
\address{Department of Physics, Emory
  University, Atlanta, GA 30322, USA }
  \author{Ilya Nemenman}
\address{Departments of Physics and Biology, Computational and Life
  Sciences Initiative, Emory University, Atlanta, GA 30322, USA}
\ead{xiang.cheng@emory.edu,ilya.nemenman@emory.edu}

\begin{abstract}
  Cells often have tens of thousands of receptors, even though only a
  few activated receptors can trigger full cellular responses.
  Reasons for the overabundance of receptors remain unclear. We
  suggest that, in certain conditions, the large number of receptors
  results in a competition among receptors to be the first to activate
  the cell.  The competition decreases the variability of the time to
  cellular activation, and hence results in a more synchronous
  activation of cells. We argue that, in simple models, this
  variability reduction does not necessarily interfere with the
  receptor specificity to ligands achieved by the kinetic proofreading
  mechanism. Thus cells can be activated accurately in time and
  specifically to certain signals. We predict the minimum number of
  receptors needed to reduce the coefficient of variation for the time
  to activation following binding of a specific ligand. Further, we
  predict the maximum number of receptors so that the kinetic
  proofreading mechanism still can improve the specificity of the
  activation. These predictions fall in line with experimentally
  reported receptor numbers for multiple systems.
\end{abstract}


\date{\today}

\section{{Introduction}}

Cellular responses to changes in the surrounding world are mediated by
protein signaling pathways. These pathways are typically activated by
modification of cell surface receptors. For example, receptors on the
surface of immune cells are a key component in detecting pathogens and
activating cellular proliferation, degranulation, and other immune
responses. The number of such receptors on the cell surface typically
is $10^4 \dots 10^6$ \cite{Coombs02,Yang10,Zaidi10}.\footnote{These
  and subsequent numbers have to be taken with a grain of salt since,
  first of all, they represent a compilation of different systems in
  different organisms and cell lines. Secondly, a variety of processes
  complicate the picture. For example, prior ligand exposure history
  may influence the number of receptors and of associated kinases,
  hence controlling the system's gain
  \cite{MacGlashan97,MacGlashan12}. Further, receptors may form
  aggregates that activate collectively, so that the ability of a
  receptor to activate a cell is a function of its physical context
  \cite{Fewtrell80,Trigoe97,Wofsy1999}. Finally, serial engagement
  allows multiple receptor (and cell) activations by the same ligand
  molecule \cite{Nag10}. Nonetheless, we believe these numbers to be
  correct within an order of magnitude, and hence illustrative. }
Surprisingly, sometimes as few as a hundred or even fewer bound
receptors are needed to fully activate a cell. For example, T-cells
can be activated by 1\dots10 receptors at the T-cell--B-cell synapse
\cite{DushekCoombs09, Valitutti10}. Rat basophilic leukemia cells
require fewer than $5\%$ active Fc$\epsilon$-RI receptors to
degranulate \cite{Ran88}.  This is not unique to immune receptors. For
example, cellular response can be triggered by only a few active
estrogen receptors \cite{Golan12}. This phenomenon of cells having an
excessive number of receptors compared to what is needed for
activation has been known for a while as a problem of "spare
receptors" \cite{Stephenson56,NICKERSON56}

Functional importance of this overabundance of receptors is not
understood, and several explanations can be considered. One
possibility is that the chemical information may not be distributed
uniformly in space, requiring receptors at multiple locations on the
cell surface. An example is the T-cell, which is commonly activated by
antigen-presenting cells through a localized synapse \cite{Lee02}.
Alternatively, the number of ligands may be very
low, and then the probability of receptor-ligand binding is increased
by having many receptors even in the well-mixed chemical kinetics
limit \cite{DushekCoombs09}. Finally, large number of receptors
facilitates clustering, promoting collective response, which may lead
to signal amplification
\cite{DeLisi81,Germain97}. 

Here we suggest that, in an alternative regime of a high ligand
concentration, the large number of receptors may play an additional
role. For example, for biological processes such as the immune
response, synchronous and short cellular activation times would allow
a concerted defense against an infection, and may be functionally
advantageous.  Similarly, concerted response is important in growth
receptor signaling. We suggest that such improvement in the accuracy
of the activation time may be achieved through a competition among
many receptors.

We envision a receptor that, when activated, produces an active
messenger molecule. A few, and maybe just one, of such molecules are
sufficient to start transcription or otherwise activate the whole
cell. Each receptor takes an extended and variable amount of time to
activate. Then the first few receptors out of many that have achieved
full activation will activate the cell. The variability of the
activation time of these fastest receptors can be smaller than that of
a typical receptor. To build an intuition, think about the
distribution of finish times for runners in a marathon. In this
analogy, each cell is a marathon with many receptors representing
runners. They all start at about the same time, launched into the
activation race by an abundant ligand, and activated receptors are
represented by runners who have finished the race. In large marathons
with thousands of participants, the time it takes to reach the finish
line for the first few runners is much smaller than the mean of the
finishing times.  More importantly, for an individual marathon, the
variability in finish times among all the runners can be fairly large,
of order of several hours. And yet the distribution of times of the
winners of many marathons has the variability of the order of minutes.
In a similar manner, the receptors that are activated first in
different cells might have activation times with small variance,
leading to a more synchronous response.

In this work, we approach the problem in the context of simplified
receptor models.  The receptor activation is first modeled as a linear
chain of $L$ states, such as ligand binding, conformational changes,
dimerization, etc.  Ligand presentation starts the progression along
the chain, and the final state represents receptor activation. With
this model, the probability distribution of time to activation of the
first receptor out of $N$ receptors, with $N\gg1$, is given by a
Weibull distribution, which has a small coefficient of variation. When
$m$ ($1\ll m\ll N$) receptors are needed to activate the cell, the
coefficient of variation decreases even further. This "racing to
activation" results in a short and less variable time to activation.

Finally, we point out that, to activate specifically to a certain
ligand, cells often use kinetic proofreading
\cite{Hopfield1974,Ninio1975,Mckeithan95}.  We show that a large
number of receptors does not necessarily interfere with the ability to
proofread. It enables cells to be activated temporally precisely, but
also specifically to certain ligands. Using biologically reasonable
parameter values, we predict the number of receptors that allows for
both of these properties.

\begin{figure}[ht!]
\centering
\includegraphics[width=22pc]{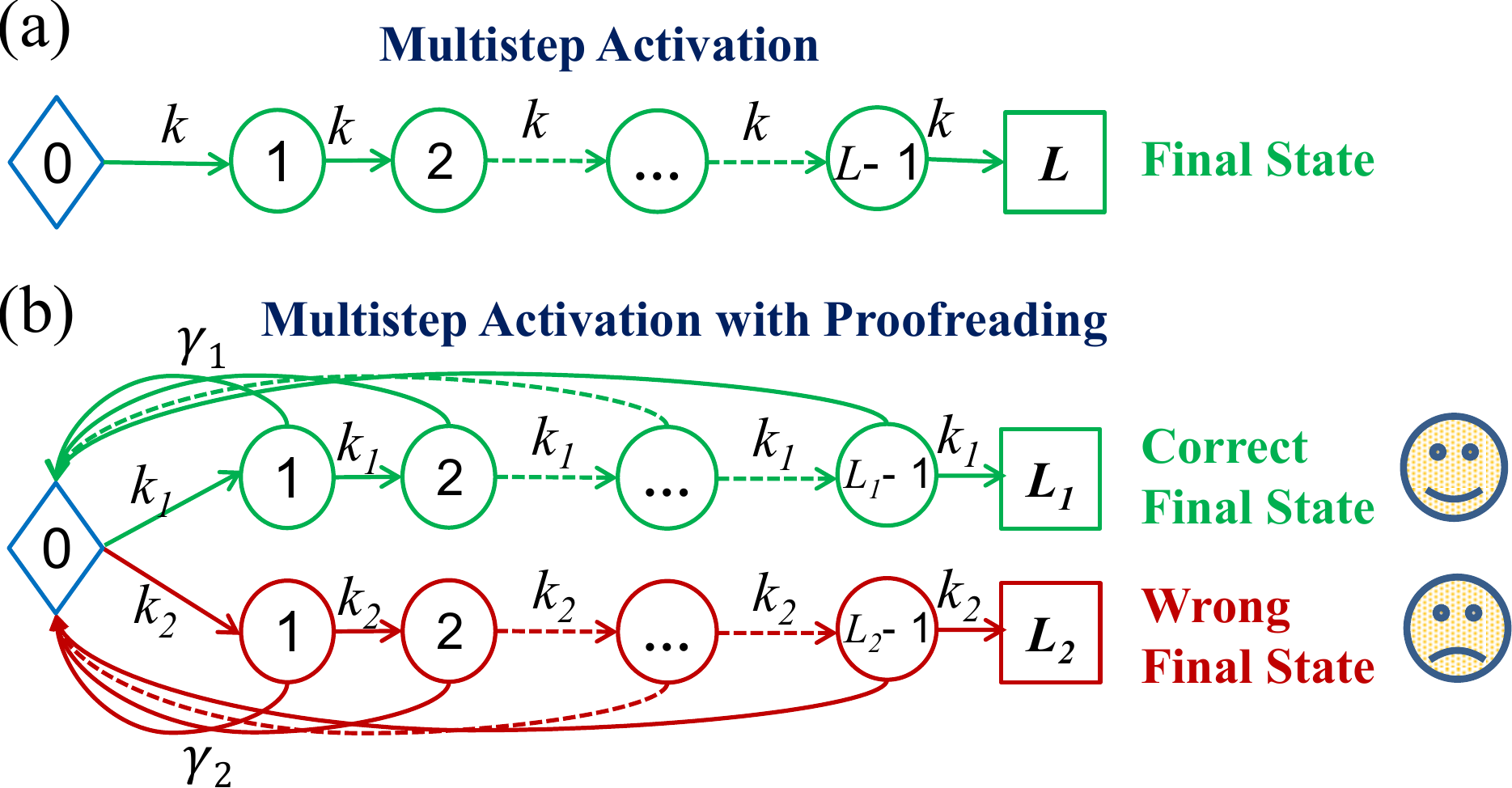}\hspace{2pc}
\caption{\label{Fig:Model}Schematic description of the model. (a) $L$
  irreversible steps with rates $k$ are needed to activate a
  receptor. (b) Kinetic proofreading, denoted by the backwards
  transitions with the rate $\gamma$, allows the receptor
  preferentially by a correct ligand.}
\end{figure}

\section{Multistep Activation}
A simplified model of receptor activation consists of a linear chain
of $L$ irreversible events where the receptor undergoes biochemical
transformations that ultimately lead to activation, see
Fig.~\ref{Fig:Model}a. In this figure, the multistep activation
process begins at the diamond (site $i=0$), and at each site the
system can transit one step towards activation with a forward rate
$k$. Reaching the right-most site, $i=L$, will lead to the receptor
activation irreversible on the time scales considered. In order to
simplify the algebraic expressions, we assume an equal forward rate
for all sites. This simplification may reduce the generality of our
results somewhat, but it does not change their dependence on $L$,
provided none of the rates $k$ are small enough to form a bottleneck.

\begin{figure}[hc]
\centering
\includegraphics[width=18pc]{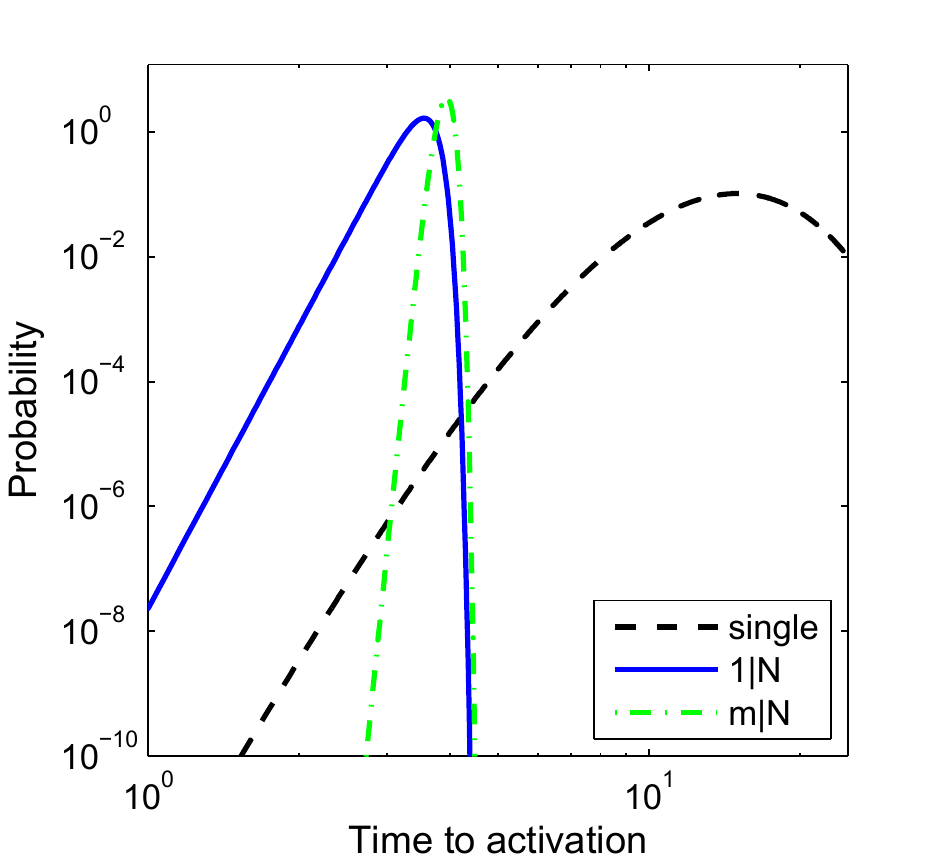}\hspace{2pc}
\caption{\label{Fig:Distributions}The probability distributions of
  time to activation with no kinetic proofreading. We use $L=16$,
  $k=1$, $N=30,000$, and $m=5$. The distribution is shown for a
  single receptor (dashed line), the first receptor out of $N$
  (solid), and the first $m$ receptors out of $N$ (dash-dotted). }
\end{figure}

The probability distribution of the time to complete each single step
in the process of the multistep activation is an exponential
distribution with rate $k$
\begin{equation}
p(t)=k\exp(-kt).
\end{equation}
The mean time is $\mu=1/k$, the variance is $\sigma^{2}=1/k^{2}$, and
the coefficient of variation is ${\rm c.v.=1}$.  For the whole process
with $L$ steps, the probability distribution of time to activation is
the well known $\Gamma$ distribution
\begin{equation}
\label{Gamma}
P_{1}(t)=\frac{k^{L}}{(L-1)!}\exp(-kt)t^{L-1}.
\end{equation}
This distribution has a broad peak. An example of the
$\Gamma$-distribution with $L=16$ and $k=1$ is shown in
Fig.~\ref{Fig:Distributions}. The moments of the $\Gamma$ distribution
are readily available:
\begin{eqnarray}
\mu&=&L/k,\\
\sigma^{2}&=&L/k^{2},\\
{\rm c.v.}&=&1/\sqrt{L}.
\label{cv1receptor}
\end{eqnarray}
The observation that multistep chemical reactions reduce the
coefficient of variation of activation time, has been well understood
computationally and experimentally, most notably in the context of
vertebrate vision \cite{DoanMendez06}.

The decrease of the coefficient of variation in proportion to
$1/\sqrt{L}$ holds true for a single receptor.  However, a cell
usually has tens of thousands of receptors, $N\gg1$. We are interested
in understanding how this large number affects the mean and the
variability of the activation times. The activation time distribution
for the first receptor out of $N$ receptors is
\begin{equation}
P_{1|N}(t)=N\ P_{1}(t)\ \left(1-\int_{0}^{t}P_{1}(t')dt'\right)^{N-1},
\end{equation}
where $P_{1}(t)$ is the $\Gamma$ distribution shown in
Eq.~(\ref{Gamma}). Here $P_1(t)$ stands for the probability of
activating the first receptor at exactly the time $t$ , and there are
$N$ choices of this receptor.  The last term represents that the other
$N-1$ receptors must not be activated before $t$. For $N\gg1$,
$P_{1|N}$ can be simplified since it is almost certain that the
first receptor is activated at the left tail of the $\Gamma$
distribution ($t\ll\frac{L-1}{k}$), although a typical receptor takes
a much longer time to be activated.  Thus, the distribution $P_{1|N}$
can be rewritten as
\begin{equation}
P_{1|N}(t)\approx\frac{L}{\alpha}\ \left(\frac{t}{\alpha}\right){}^{L-1}\ \exp\left[-\left(\frac{t}{\alpha}\right)^{L}\right].
\end{equation}
where $\alpha=\left(\frac{L!}{Nk^{L}}\right)^{\frac{1}{L}}$. $P_{1|N}$
is the Weibull distribution, which we illustrate in
Fig.~\ref{Fig:Distributions}. The Weibull distribution has the
following well-known statistical properties
\begin{eqnarray}
\mu&=&\alpha\Gamma\left(1+\frac{1}{L}\right),\\
\sigma^{2}&=&\alpha^{2}\left[\Gamma\left(1+\frac{1}{L}\right)-\Gamma^{2}\left(1+\frac{1}{L}\right)\right],\\
{\rm c.v.}&=&\frac{\sqrt{\psi^{(1)}(1)}}{L}\approx \frac{1.28}{L},
\end{eqnarray}
where $\Gamma(x)$ is the $\Gamma$ function and $\psi^{(1)}(x)$ is a
polygamma function defined as $\psi^{(1)}(x)=\frac{{\rm d}^2}{{\rm
    d}x^2}\ln\Gamma(x)$.  The coefficient of variation goes as
$\propto\frac{1}{L}$, which decreases much faster with $L$ than it
does for a single receptor activation, $\frac{1}{\sqrt{L}}$.

One can envision a situation where a cooperative action of a few
receptors, or more precisely of their activated messages, is needed to
activate the whole cell.  We can extend our model by studying the
distribution of time to activating $m$ out of $N$ receptors,
$P_{m|N}(t)$, with $1\ll m\ll N$. There are $N \choose{m}$
possibilities to choose the set of $m$ receptors to be activated out
of total $N$ receptors. We require that one of the $m$ receptors
activates at exactly time $t$, which has a probability of $m
P_{1}(t)$. The other $m-1$ receptors will finish the process at some
point before time $t$ which is described by
$\left[\int_{0}^{t}P_{1}(t')dt'\right]^{m-1}$. The rest of the $N-m$
receptors are not activated up to time $t$. Combining the terms, we
get
\begin{equation}
\fl
P_{m|N}(t)=\frac{N!}{m!\ (N-m)!} m\ P_{1}(t) \left[\int_{0}^{t}P_{1}(t')dt'\right]^{m-1}\ \left[1-\int_{0}^{t}P_{1}(t')dt'\right]^{N-m}.
\end{equation}
With the same approximation $t\ll\frac{L-1}{k}$, $P_{m|N}(t)$ can be rewritten as
\begin{equation}
  P_{m|N}(t)\approx\frac{N!}{m!\ (N-m)!}  m\ P_{1}(t)\left(\frac{k^{L}}{L!}\right)^{m}\ t^{mL-1}\ \exp\left(-\frac{Nk^{L}}{L!}t^{L}\right).
\end{equation}
This distribution is shown in Fig.~\ref{Fig:Distributions} with the
usual parameter values, and $m=5$. Activating multiple receptors has a
narrower distribution with a smaller coefficient of variation. We
calculate
\begin{eqnarray}
  \mu\approx\frac{1}{(m-1)!}\left(\frac{L!}{N}\right)^{\frac{1}{L}}\Gamma\left(m+\frac{1}{L}\right)\frac{1}{k},\\
  \sigma^{2}\approx\frac{1}{(m-1)!}\left(\frac{L!}{N}\right)^{\frac{2}{L}}\Gamma\left(m+\frac{2}{L}\right)\frac{1}{k^{2}},\\
  {\rm c.v.}\approx\frac{\sqrt{\psi^{(1)}(m)}}{L}\approx\frac{1}{L\sqrt{m}},
\end{eqnarray}
where the last approximation comes from the first few terms in the
asymptotic expansion of
$\psi^{1}(m)\approx{1}/{m}-{1}/{2m^2}+{1}/{24m^3}$, assuming
$m\gg1$. Thus the coefficient of variation for cooperative activation
is further reduced compared to the activation by the first
receptor. 

The results of the three discussed cases (single receptor, the first
out of many receptors, and the first $m$ out of many receptors needed
for activation) are summarized in Table \ref{Table:NoKPR}. The racing
to activation mechanism with a large number of participating receptors
can reduce the coefficient of variation of the activation time to
below $5\%$ for reasonable parameters. This could mean a more
synchronous activation of cells.
\begin{center}
\begin{table}
\begin{tabular}{|c|c|c|c|c|}
  \hline 
  & Mean time & Variance & ${\rm c.v.}$ & example of ${\rm c.v.}$\tabularnewline
  \hline 
  \hline 
  1 receptor & $\frac{L}{k}$ & $\frac{L}{k^{2}}$ & $\frac{1}{\sqrt{L}}$ & 25.0\%\tabularnewline
  \hline 
  $1|N$ receptors & $\alpha\Gamma(1+\frac{1}{L})$ & $\alpha^{2}[\Gamma(1+\frac{2}{L})-\Gamma^{2}(1+\frac{1}{L})]$ & $\sim\frac{1.28}{L}$ & 8.0\%\tabularnewline
  \hline 
  $m|N$ receptors & $\frac{1}{(m-1)!}[\frac{L!}{N}]^{\frac{1}{L}}\Gamma(m+\frac{1}{L})\ \frac{1}{k}$ & $\frac{1}{(m-1)!}[\frac{L!}{N}]^{\frac{2}{L}}\Gamma(m+\frac{2}{L})\frac{1}{k^{2}}$ & $\sim\frac{1}{L\sqrt{m}}$ & 2.8\%\tabularnewline
  \hline 
\end{tabular}
\caption{Comparison of activation time distribution for three
  different activation processes.
  The example uses the values $L=16,\ k=1,\ m=5,$ and $N=30,000$}
\label{Table:NoKPR}
\end{table}
\end{center}

\section{Multistep activation with kinetic proofreading}
Receptors can bind many non-specific ligands, and they employ {\em
  kinetic proofreading} to increase specificity and activate
predominantly following binding of a specific ligand
\cite{Mckeithan95,Goldstein08}. Our model of kinetic proofreading
follows Ref.~\cite{Munsky09}, and is shown in
Fig.~\ref{Fig:Model}b. We assume that there are two branches of
sequential events in the receptor state diagram: a correct branch and
a wrong one. The two branches represent activation sequences following
binding of a specific/nonspecific ligand molecules, with rates $k_1$
and $k_2$ for the correct/wrong branches, respectively. For both
branches, kinetic proofreading is represented by possible jumps back
to the initial state with rates $\gamma_{1}$ and $\gamma_2$,
respectively.  These jumps can correspond to ligand unbinding from the
receptor, followed by return of the receptor into the initial, fully
inactive state. If $\gamma_2\gg\gamma_1$ (that is, a nonspecific
ligand unbinds faster than the specific one), then the receptor has a
much lower probability to reach the final, absorbing state along the
wrong branch than along the correct one, hence increasing the
specificity. While this model is crude, it nonetheless captures the
basic physical properties of the kinetic proofreading process.

\begin{figure}
\centering
\includegraphics[width=0.5\textwidth]{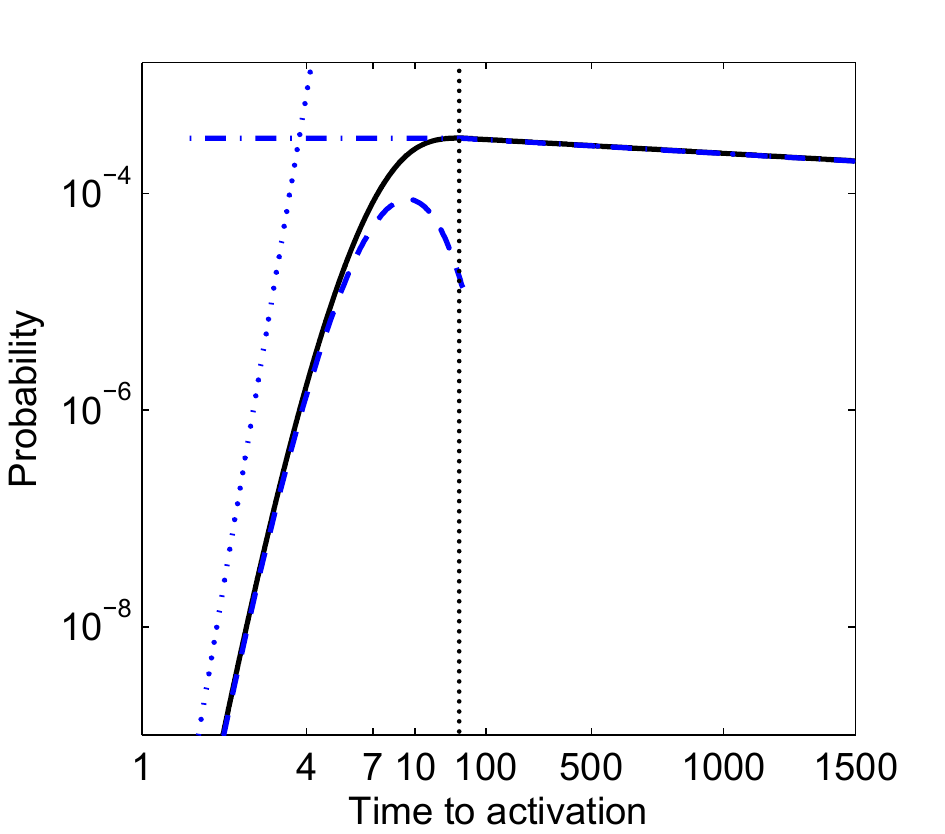}
\caption{\label{Fig:KPRdistr}The probability distribution of time to
  activation with kinetic proofreading for a single branch with
  $L=16$, $k=1$, and $\gamma=0.6$. The vertical dotted line is the $t$
  of maximum probability.  On left side of this line, the curves are
  in log-log scale; while the curves on the right are in log-linear
  scale. Solid line comes from a numerical solution of the master
  equation describing the receptor. We see the power law distribution
  when $t$ is small, and the exponential distribution when $t$ is
  large. These are illustrated with the right tail exponential
  asymptotic, Eq.~(\ref{Eq:KPR2}) (dash-dotted line); the left tail
  asymptotic, Eq.~(\ref{lefteq}) (dashed line), and the power law line
  $k^L t^{L-1} /(L-1)!$ (dotted line). }
\end{figure}

Kinetic proofreading typically leads to a near-exponential
distribution of completion times \cite{Munsky09,Bel10}, 
which has ${\rm c.v.}\approx1$. The distribution is
different from the exponential only over a very short initial time
period since it is difficult to traverse the sequence of $L$ steps to
activation quickly, see Fig.~\ref{Fig:KPRdistr}. This has a potential
of interfering with the racing to activation mechanism for improving
temporal precision. Indeed, a receptor with an exponentially
distributed activation time is equivalent to a single-step receptor,
$L=1$. In this case the corresponding $\Gamma$ and the Weibull
distributions are exponential as well, providing no reduction in the
coefficient of variation.

However, if the number of receptors is sufficiently large, then the
first few receptors will have a substantial chance to activate during
the initial, nonexponential phase of the completion time
distribution. That is, they will activate quickly in a linear sequence
of events, as in the previous Section, having no time to revert back
to the initial state through the proofreading mechanism. Thus if the
number of receptors is sufficient to complete so quickly along the
correct branch, the temporal precision will increase through the
racing to activation mechanism. On the other hand, since $\gamma_2$ is
larger, the receptor trying to activate over the same time along the
wrong branch may have sufficient time to revert to the initial state
repeatedly, falling into the exponential part of the completion time
distribution. This suggests that the number of receptors on the cell
surface should also be limited from above, so that wrong activation
has a low probability of happening in the pre-exponential part of the
activation distribution.

Mathematically, we can summarize these arguments as follows. Following
Ref.~\cite{Bel10}, we approximate the probability distribution of
completing along a single branch of the state space as:
\begin{eqnarray}
  P(t)&\approx&\frac{k^{L}}{(L-1)!}t^{L-1}\exp\left[-\left(\gamma+k\right)t\right],\
  \ \ t\ll t_{C},\label{lefteq}\\
    P(t)&\approx&\frac{1}{\mu}\exp\left(-\frac{t}{\mu}\right),\ \ \ \ \ \ \ \ \ \ \ \  \ \ \  \  \ \ \ \ \ \ \  \ t\gg t_{C}.
\label{Eq:KPR2}
\end{eqnarray}
Here the short-time limit starts as the $\Gamma$-distribution, and
there is an additional decrease in probability of completing at time
$t$ by $\exp(-\gamma t)$ because of kinetic proofreading. This
approximation is valid to the left of its maximum,
\begin{equation}
  t_{C}=\frac{L-1}{k+\gamma}.
\end{equation}
Further, $\mu$ is the mean activation time, calculated in Ref.~\cite{Bel10} to
be
\begin{equation}
\mu=\frac{1}{\gamma}\left[\left(1+\frac{\gamma}{k}\right)^L-1\right].
\end{equation}
The approximation is illustrated in Fig.~\ref{Fig:KPRdistr}.

The variance reduction along the correct branch requires that the
first receptors are activated before $t_{C_1}$, that is:
\begin{equation}
  N\  \intop_{0}^{t_{C_1}}\frac{k_1^{L_1}}{(L_1-1)!}t^{L_1-1}\exp\left[-\left(k_1+\gamma_1\right)t\right]dt\gg
  1,
\label{eq:bnd1}
\end{equation}
where, as before, the subscript $1$ stands for the correct branch.
Using the standard expressions for the incomplete $\Gamma$-function,
we thus derive the minimum number $N_{\min}$ of receptors needed for
variability reduction along the correct branch
\begin{equation}
 N\gg N_{\rm min}=\frac{\left(1+\frac{\gamma_1}{k_1}\right)^{L_1}}{1- \exp\left(1-L_1\right)\ \displaystyle \left(\sum_{i=0}^{L_1-1}\frac{(L_1-1)^{i}}{i!}\right)}.  
\end{equation}

This sets the lower bound on the number of receptors on the cell
surface. However, if the number of receptors was too high, some of the
ones bound by incorrect ligands would activate in the left tail of the
distribution, before the kinetic proofreading has a chance to kick
in. The requirement that the probability of early activation over the
wrong branch among all receptors is negligible sets the upper limit on
the receptor number:
\begin{equation}
  N\
  \intop_{0}^{t_{C_2}}\frac{k_2^{L_2}}{(L_2-1)!}t^{L_2-1}\exp\left[-\left(k_2+\gamma_2\right)t\right]dt\ll 1,
\end{equation}
where the subscript $2$ stands for the wrong branch.  This results in 
\begin{equation}
  N\ll N_{\rm max}= \frac{ \left(1+\frac{\gamma_2}{k_2}\right)^{L_2}}{1-\exp\left(1-L_2\right)\ \displaystyle \left(\sum_{i=0}^{L_2-1}\frac{(L_2-1)^{i}}{i!}\right)}. 
\end{equation}

While we do not expect that the bounds derived from our manifestly
oversimplified model predict the real number of receptors on the cell
surface, it is still worthwhile to verify if the obtained bounds are
meaningful, so that kinetic proofreading over a wrong branch can be
consistent with narrow activation time over the correct one.  As in
Ref.~\cite{Bel10}, we choose $L=16$, and $k_1=k_2=1$, the same for
both the correct and the incorrect branches. To choose the kinetic
proofreading rates $\gamma_{1,2}$, we note that they correspond, for
example, to unbinding of a ligand from the receptor complex, so that
the log-ratio of the rates are related to the difference of the
binding free energies $\Delta G_{1,2}$ of the correct and the
incorrect ligands:
\begin{equation}
\Delta G_1-\Delta G_2=k_{B}T\ln\frac{\gamma_2}{\gamma_1}.
\end{equation}
Assuming the difference in the binding free energies of a few thermal
energies, we get
\begin{equation}
\frac{\gamma_2}{\gamma_1}\sim2\dots10.
\end{equation}
Now choosing $\gamma_1\approx0.7$, we get the minimum number of
receptors $N_{\rm min}\sim 10^{4}$, and the maximum number $N_{\rm
  max}\sim 10^7$ or more. The window between $N_{\min}$ and $N_{\max}$
is sufficiently large to allow the choice of receptor number that
satisfies both the proofreading and the narrow activation
time. Further, these numbers are realistic in the context of
biological receptor numbers, which range between $10^4$ and $10^6$.

\section{Conclusion}

Biological systems must operate under constraints posed by the
physical world. One of such constraints is robustness to intrinsic
noise that comes about from the small number of stochastically
activated molecular components in cellular networks. Performing
averages of different kinds (over time, space, or different molecular
species) is, essentially, the only way of ensuring such robustness. In
this context, it has been understood for many years that sequential
activation of $L$ molecular components averages the times of the
activation steps and reduces the coefficient of variation of the time
to activation, which scales then as $\sim1/\sqrt{L}$
\cite{rieke98}. This has been observed experimentally in the context
of vertebrate phototransduction \cite{DoanMendez06}. Here we propose a
mechanism that can further reduce the variability to $\sim 1/L$, and
even $\sim 1/L\sqrt{m}$ (where $m$ is the cooperativity of the
activation process), which may be a substantial improvement for long
activation sequences and large cooperativities. The reduction of
variability is achieved by having many equivalent multi-step entities,
competing to be the first one to activate the whole system. We believe
that this mechanism, which we call {\em racing to activation}, has not
been reported previously in the literature.

Our focus here is on the general noise suppression mechanism, and not
on any specific biological system. Nonetheless, it is clear that
various receptor-mediated signaling systems are likely to be examples
of where this mechanism will be applicable. The fact that receptors
exist on cell surfaces in very large numbers, far in excess of what is
needed to fully activate a cell, is encouraging. However, the large
number of receptors needed for reduction of the temporal noise in the
racing to activation mechanism can come in conflict with the
specificity of signaling.  It is, therefore, reassuring that, within
our model, there is a range of receptor numbers that allows cells to
maintain specificity to certain ligands, and yet be activated
precisely in time.

Another possible application of our model is in development. There a
crucial question is the precision of position determination afforded
by morphogen gradients
\cite{Tostevin:2007hh,Gregor:2007du,Lander:2009jc}. While most
analyses are done for steady state gradients, cells may need to make
developmental commitments based on pre-steady state transients
\cite{Saunders:2009gp}. Adding reverse reactions to
Fig.~\ref{Fig:Model}(a), with rates $k_{\rm reverse}<k$, our model can
be interpreted as a biased (nonequilibrium) random walk of a morphogen
particle in the physical space, where the first $m$ of such particles
that reach a morphogen detector at $L$ activate it. Our arguments
suggest that such temporal, first-passage triggering of developmental
commitments can be made very precise.

While too coarse for specific details, our results can be used to make
general, verifiable predictions about relations between the number of
receptors, the similarity of specific and non-specific ligands, the
cooperativity of receptor action on the one hand, and the variability
of the activation time on the other. Such data, when available, will
shed light on whether the proposed mechanism is used in cellular
processes, and to which extent.

\ack IN would like to thank Brian Munsky and Nicolas Hengartner for
stimulating discussions during early stages of this work and Byron
Goldstein for his help in understanding a proper immune signaling
context for our model. The authors are further grateful to attendees
of The Sixth q-bio Conference, who provided us with invaluable
comments. This work was funded in part by the James S.\ McDonnell
foundation Grant No.\ 220020321.

\section*{References}
\bibliographystyle{unsrt}

\end{document}